\documentclass[iicol]{sn-jnl}

\usepackage{graphicx}%
\usepackage{multirow}%
\usepackage{amsmath,amssymb,amsfonts}%
\usepackage{amsthm}%
\usepackage{mathrsfs}%
\usepackage[title]{appendix}%
\usepackage{xcolor}%
\usepackage{colortbl}
\usepackage{textcomp}%
\usepackage{manyfoot}%
\usepackage{nicematrix} 
\usepackage{caption}
\usepackage[normalem]{ulem}

\usepackage{tabularray}
\usepackage{booktabs}

\usepackage{natbib}
\setcitestyle{authoryear,round} 

\usepackage[ruled]{algorithm2e}
\usepackage{dsfont}%
\usepackage{listings}%
\usepackage{physics}
\usepackage{tikz}
\usetikzlibrary{fit,calc}
\usetikzlibrary{quantikz}
\usepackage{subfigure}

\renewcommand{\figurename}{Figure}

\definecolor{erblue}{HTML}{0082F0}
\definecolor{erorange}{HTML}{FF8C0A}
\definecolor{ergreen}{HTML}{0FC373}
\definecolor{erpurple}{HTML}{AF78D2}
\definecolor{eryellow}{HTML}{FAD22D}
\definecolor{erred}{HTML}{FF3232}

\def\BibTeX{{\rm B\kern-.05em{\sc i\kern-.025em b}\kern-.08em
    T\kern-.1667em\lower.7ex\hbox{E}\kern-.125emX}}

\theoremstyle{thmstyleone}%

\theoremstyle{thmstyletwo}%

\theoremstyle{thmstylethree}%

\begin{document}

\colorlet{pink}{red!40}
\colorlet{green}{green!60}

\newcommand{\red}[1]{{\color{red}#1}}

\newcommand{\bb}[1]{{\color{erblue}Bence: #1}}
\newcommand{\zz}[1]{{\color{purple}Cimbi: #1}}
\newcommand{\nd}[1]{{\color{erred}Dani: #1}}
\newcommand{\zk}[1]{{\color{ergreen}Zs\'ofi: #1}}
\newcommand{\cc}[1]{{\color{teal}Csaba: #1}}
\newcommand{\hp}[1]{{\color{erorange}HPeti: #1}}

\title[Article Title]{Hybrid Quantum-Classical Reinforcement Learning in Latent Observation Spaces
}

\author[1,2,3]{\fnm{Dániel} \sur{T. R. Nagy}} 

\author[1,2,3]{\fnm{Csaba} \sur{Czabán}} 

\author[1,2,3]{\fnm{Bence} \sur{Bakó}} 

\author[3]{\fnm{Péter} \sur{Hága}} 

\author[3]{\fnm{Zsófia} \sur{Kallus}} 

\author[1,2,4]{\fnm{Zoltán} \sur{Zimborás}} 

\affil[1]{\centering\orgdiv{Eötvös Loránd University}, \city{Budapest}, \country{Hungary}} 

\affil[2]{\orgdiv{HUN-REN Wigner Research Centre for Physics}, \city{Budapest}, \country{Hungary}} 

\affil[3]{\orgdiv{Ericsson Research}, \city{Budapest}, \country{Hungary}}

\affil[4]{\orgdiv{Algorithmiq Ltd.}, \city{Helsinki}, \country{Finland}}
\affil[]{\\ Email: nagy.dani@wigner.hun-ren.hu}

\abstract{
Recent progress in quantum machine learning has sparked interest in using quantum methods to tackle classical control problems via quantum reinforcement learning.
However, the classical reinforcement learning environments often scale to high dimensional problem spaces, which represents a challenge for the limited and costly resources available for quantum agent implementations.
We propose to solve this dimensionality challenge by a classical autoencoder and a quantum agent together, where a compressed representation of observations is jointly learned in a hybrid training loop. The latent representation of such an autoencoder will serve as a tailored observation space best suited for both the control problem and the QPU architecture, aligning with the agent's requirements.
A series of numerical experiments are designed for a performance analysis of the latent-space learning method. Results are presented for different control problems and for both photonic (continuous-variable) and qubit-based agents, to show how the QNN learning process is improved by the joint training.
}


\maketitle

\section{Introduction}\label{sec1}

\par Machine learning (ML) solutions are rapidly adapted to many complex control applications requiring the modeling power of deep neural networks, e.g., of reinforcement learning (RL) agents.
However, many of these application fields present problems where deep learning struggles with the complexity of hidden patterns leading to bottlenecks in performance and efficiency \citep{ai-complexity-1,ai-limits-1}. Thus, using accelerators and emerging technologies for ML efficiency improvement is an active field of research \citep{google-tpu-v4,ai-accelerator-survey}.

One of these emerging technologies is quantum computing that can be adapted to classical problem spaces, acting as an accelerator. This approach is a rapidly growing sub-field of quantum machine learning (QML) \citep{huang2022quantumAdvantage, xiao2023practicalQAdvantage,gyurik2022towardsQAdvantage,riste2017demonstrationQAdvantage}.

Today, QML approaches can only be efficiently used as accelerators as part of hybrid solutions where the problem is jointly solved by quantum and classical parts.

As a part of this research area, Quantum Reinforcement Learning (QRL) is focusing on quantum algorithms for reinforcement learning models leveraging quantum computational resources \citep{kwak-qrl, meyer2022survey}. 
Current quantum processing units are however still limited in their scale and error tolerance \citep{bharti2022noisy,lau2022nisq}, therefore developing and testing heuristic QRL approaches with real-world applications remains a challenge. A key issue in many  QRL methods is that several real-world problems require encoding high-dimensional feature vectors into the QPU's initial quantum state.

To tackle the high dimensionality of feature vectors, classical deep learning pipelines often use Autoencoders (AEs) \citep{Hinton_AE, Goodfellow-et-al-2016} or Variational Autoencoders (VAEs) \citep{KingmaVAE,KingmaVAE_Introduction} to  extract relevant low-dimensional features from high-dimensional complex problem spaces, and perform computations on the latent-space vectors. 
Prominent examples of this approach are latent diffusion models, in which images are compressed by using VAEs and generative modeling is performed on latent images \citep{latentDiffusion}.
AEs have also been successfully used in reinforcement learning tasks, e.g., the Dreamer model is used for learning a latent-space representations with a VAE and for policy optimization by latent trajectory imagination \citep{Lange2010DeepAN,vanhoof,dreamer}.

QRL agents have been recently used to solve standard baseline problems in classical control benchmark environments of both discrete and continuous action spaces \citep{Dunjko_AdvancesinQRL,Lamata_photonicQRL,Chen_TN_AE_QRL,WuQRL_continuous,dnagy}.
Furthermore, QRL has been tested in various, e.g., free-energy-based RL with quantum Boltzmann machines \citep{Schenk_2024_QRL_FOR_CERN}, QRL with various dimensionality reduction methods \citep{DimRedQRL}, evolutionary learning strategies using tensor network-variational quantum circuits \citep{Chen_TN_AE_QRL}, photonic QRL with photonic proximal policy optimization (PPO) using on manual feature selection \citep{dnagy}.

\par
In this paper, we present a method to train AEs and quantum agents jointly in a hybrid quantum-classical training loop such that the AE learns a particular feature extraction best suited for both the given problem and the QPU architecture. 

\par
The outline of this paper is as follows: in Section \ref{sec:background}, we briefly review the ML and QML background of our work. Section \ref{sec:latent-qrl} presents our novel method of training variational quantum agents based on parametrized quantum circuits (PQC) in latent observation spaces by using AEs and a combined loss function. In Section \ref{sec:numexp-setup}, we present the setup of our numerical experiments, while Section \ref{sec:numexp-results} discusses the obtained numerical results. Finally, in Section \ref{sec:conclusion} we derive our conclusions and present some outlook for future work.

\section{Background}
\label{sec:background}
In this section, we briefly review the technical background of our work, specifically, we review classical machine learning techniques of autoencoders and reinforcement learning, and also present the relevant methods of quantum reinforcement learning.

\subsection{Autoencoders}
\label{sec:ae-background}
Popular algorithmic feature reduction techniques such as Principal Component Analysis (PCA), Independent Component Analysis (ICA) or  t-Distributed Stochastic Neighbor Embedding (t-SNE) are often unable to find the best low-dimensional features for high-complexity data, e.g., as visual data. To overcome this, deep neural network based feature reduction techniques such as AEs and VAEs have been implemented and are widely used to learn efficient dimensionality reductions on complex datasets, as well as mitigating or correcting specific noise and error patterns \citep{VincentDenoisingAE, YasenkoAEDenoising,  RestrepoFeatureVAE}.

AEs \citep{Hinton_AE, Goodfellow-et-al-2016} are a type of artificial neural network used in unsupervised machine learning and deep learning, primarily for data compression, feature learning, and dimensionality reduction. The key idea behind AEs is to learn a compact representation (encoding) of input data by training the network to reconstruct its input as accurately as possible. AEs consist of an encoder $\mathcal{E} : \mathbb R^D \mapsto \mathbb R^L$, and a decoder  
$\mathcal{D} : \mathbb R^L \mapsto \mathbb R^D$. 
The encoder transforms data points $\mathbf{x} \in \mathbb R^D$ into a \textit{latent space} representation $\mathbf{z} = \mathcal{E}(\mathbf{x}) \in \mathbb R^L$, whereas the decoder tries to reconstruct it: $\hat{\mathbf{x}} = \mathcal{D}(\mathbf{z})$. When $L<D$, the encoder produces a compact 
representation of the data. 
$\mathcal{E}$ and $\mathcal{D}$ are typically deep neural networks trained jointly via gradient descent optimization to minimize the reconstruction error or loss $\mathcal{L}(\mathbf{x}, \mathcal{D}(\mathcal{E}(\mathbf{x}))$.
The loss function $\mathcal{L}(\mathbf{x}, \mathcal{D}(\mathcal{E}(\mathbf{x}))$ can be as simple as a mean-squared error (MSE) for continuous data \citep{Goodfellow-et-al-2016}, categorical cross-entropy for categorical data \citep{Goodfellow-et-al-2016}, or something more complex like Fréchet inception distance (FID) for visual data \citep{yu2021frechet}.

\subsection{Classical reinforcement learning}
\label{sec:rl-background}

Reinforcement Learning (RL) is a sub-field of machine learning aiming to train RL-\textit{agent}s such that these select the optimal \textit{action} to be performed in an \textit{environment}. 
Optimal actions maximize the cumulative \textit{reward}s collected throughout the agents lifecycle. 
RL algorithms are either model-based, i.e., learning a predictive model of the environment, or model-free, i.e., learning a control policy directly without learning the underlying environment model. 

Policy-based RL methods use a policy network $\pi_{\theta}$, which defines an action probability distribution $\pi_{\theta}(\cdot|\mathbf{s}_t)$ over the space of allowed actions for every state $\mathbf{s}_t$ of the environment. 

At each timestep an action $\mathbf{a}_t $ is sampled from the distribution $\pi_{\theta}(\cdot|\mathbf{s}_t)$ and executed in the environment, which in turn gives a scalar reward $r_t$ as feedback. In general, a \textit{trajectory} is defined as a sequence of states and actions $\tau = \{\mathbf{s}_1, \mathbf{a}_1, ... \mathbf{s}_T, \mathbf{a}_T\}$, and the corresponding trajectory reward is 
\begin{equation}
    R(\tau) = \sum\limits_{t=1} ^ \infty \gamma^t r_t \approx 
    \sum\limits_{t=1} ^ T \gamma^t r_t,
\end{equation}
with $\gamma \in (0, 1)$ being the discount factor. The general goal of RL is to find the optimal policy $\pi^{*}_{\theta}$ which maximizes the expected trajectory reward:
\begin{equation}
    \pi^{*}_{\theta} = \underset{\pi_\theta}{\arg\max}\,\,
    \mathbb{E}_{\tau\sim\pi_\theta} R(\tau) .
\end{equation}
To find the optimal policy, one usually estimates the expected return $\mathbb{E}_{\tau\sim\pi_\theta} R(\tau)$ and uses this to calculate the gradients $\nabla_{\theta}\mathbb{E}_{\tau\sim\pi_\theta} R(\tau)$ and perform gradient ascent optimization. This method is called the vanilla policy gradient (VPG) method \citep{VPGPaper}, which suffers from instability \citep{TRPOPaper,PPOPaper}, and hence other algorithms like proximal policy optimization (PPO) were invented.

\begin{figure*}[h]
    \centering
    \includegraphics[height=8cm]{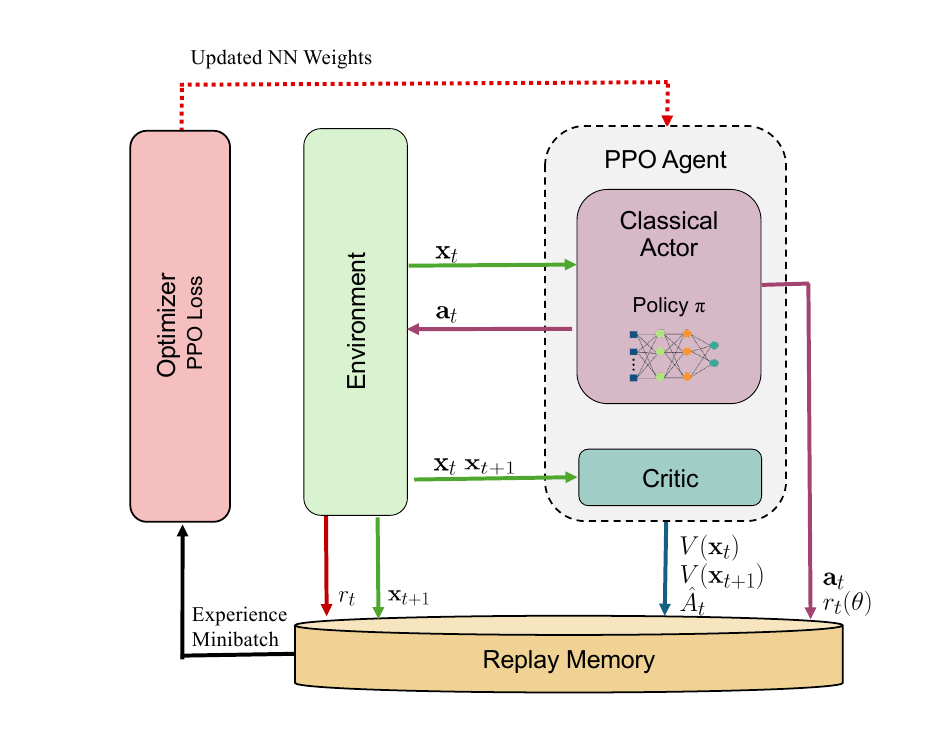}
    \caption{\textbf{Classical PPO algorithm training loop.} The PPO agent is composed of two neural networks: actor (a.k.a. policy) and critic. The policy network receives an observable from the environment and decides the actions to be performed. The critic network receives state vectors and computes the state values as well as the action advantages. Rewards, states, actions, values, advantages and policy ratios are stored in a replay memory, from which minibatches are sampled during training.}
    \label{fig:classical-ppo-loop}
\end{figure*}

Using the notation of \citep{PPOPaper}, here we briefly summarize the PPO algorithm. PPO is an RL algorithm designed to overcome the instability of VPG, by introducing a clipped surrogate objective. In order to do so, PPO calculates the probability ratios $r_t(\theta)=\pi_{\theta}(\mathbf{a}|\mathbf{s})/\pi_{\theta_{\textrm{old}}}(\mathbf{a}|\mathbf{s})$, and clips it outside the proximity region defined by $\epsilon$ as:
\begin{equation*}
    \textrm{clip}(r_t(\theta), \epsilon) = 
    \begin{cases}
    1-\epsilon, & \textrm{if}~r_t(\theta) < 1-\epsilon \\
    r_t(\theta), & \textrm{if}~r_t(\theta) \in[1-\epsilon, 1+\epsilon] \\
    1+\epsilon, & \textrm{if}~r_t(\theta) > 1+\epsilon
    \end{cases}
\end{equation*}
Furthermore, an \textit{advantage} estimate $\hat A_t$ is calculated which represents the advantage of the updated policy compared to the old policy. The clipped surrogate loss can be calculated as 
\begin{equation*}
\hspace{-0.2cm}
    \mathcal{L}^{\text{CLIP}}(\theta) = \mathbb{E}_t \left[ \min\left( r_t(\theta)\hat A_t, \textrm{clip}\left(r_t(\theta),\epsilon\right)\hat A_t\right) \right],
\end{equation*}
where the advantages $\hat A_t$ are calculated via generalized advantage estimation \citep{GAEPaper}:
\begin{equation}
    \hat A_t = \sum\limits_{l=0}^{T-t-1} (\gamma\lambda)^{l}\delta_{t+l},
\end{equation}
and $\delta_t = r_t + \gamma V^{\pi}(\mathbf{s}_{t+1}) - V^{\pi}(\mathbf{s}_t)$. The value function $V^{\pi}(\mathbf{s}_t)$ is a neural network separate from $\pi_\theta$ and is usually trained via mean-squared error loss:
\begin{equation}
    \mathcal{L}^{VF} = \underset{t}{\mathbb E} \left[\left(V^{\pi}(\mathbf{s}_t) - V_{\textrm{targ}}^{\pi}(\mathbf{s}_t)\right)^2\right].
\end{equation}
The clipped surrogate objective $L^{\text{CLIP}}$ ensures that the agent does not suffer from destructively large updates during training. This however limits the exploration, as the agent might not be able to explore better policies and solve the problem. To overcome the exploration issue, it is straightforward to incentivize exploration by adding a positive entropy term to the loss:
\begin{equation}
\hspace{-0.2cm}
    \mathcal{L}^{\text{PPO}} = -\left[ \mathcal{L}^{\text{CLIP}}(\theta) +
 c_1 S \left[ \pi_{\theta} \right] + c_2\text{Reg}(\theta) \right],
 \label{eq:ppofull}
\end{equation}
where $\text{Reg}(\theta)$ is some regularization term, usually $L^2$ regularization, and $S \left[ \pi_{\theta} \right]$ is the entropy of the policy.
Note that we want to maximize the PPO objective via gradient ascent, however common AI software packages such as tensorflow perform gradient descent by default, hence the minus sign of the objective. Figure \ref{fig:classical-ppo-loop} summarizes the components of this complex classical PPO training loop.

\subsection{Quantum reinforcement learning}
\label{sec:qrl-background}

Quantum Reinforcement Learning (QRL) is an emerging area of quantum computing and reinforcement learning that explores how quantum algorithms and quantum resources can be applied to enhance traditional reinforcement learning techniques \citep{saggio2021experimental, meyer2022survey}. 
The state space of quantum circuits scales exponentially with the number of qubits/qumodes and hence quantum approaches potentially enable the solutions of machine learning problems that are intractable for classical computers.

Many QRL approaches are based on quantum neural networks (QNNs) in a hybrid quantum-classical training loop \citep{dnagy,kwak-qrl,Dunjko_AdvancesinQRL,Lamata_photonicQRL,WuQRL_continuous,Chen_TN_AE_QRL, vqc_qrl_rodrigo}, other QRL approaches use Grover's search algorithm \citep{qrl_grover,li2020quantum_grover}, quantum annealing \citep{qrl_annealing} or other quantum algorithms \citep{sannia2023hybrid} to optimize agents. QNN-based RL agents use parametrized quantum circuits (PQCs) as their core component and optimize PQC parameters via classical algorithms such as gradient descent to reach an optimal reward score in a specific environment.
In this case, the parameters of the PQC are stored in the memory of a classical device and the PQC is used to generate a quantum state that is measured to obtain an action as its output. The classical optimizer computes the PPO loss, from which one can compute the gradients of the PQC parameters using the quantum device. Once the PQC gradients are calculated, the classical optimizer can compute the updated gate parameters for each quantum gate of the circuit and continue the training loop with the updated circuit. 

Formally, a quantum neural network is a parametrized unitary operator $U(\pmb\theta)$ defined by a sequence of layers composed of quantum gates. This unitary is acting on the initial quantum state of the system, $\rho_i$, producing a final state $\rho_f = U(\pmb\theta) \rho_i U^\dagger(\pmb\theta)$. Subsequently, measurements are performed on $\rho_f$ to obtain the QNN outputs. In the case of QNN-based RL agents, we need to embed the state $\mathbf{x}$ of the system into the PQC. Assuming that each layer of the QNN has trainable parameters $\pmb\theta_l$, we can define an encoding unitary $U_E(\mathbf{x})$ which encodes classical features. The total unitary applied to the initial state $\rho_i$ is therefore
\begin{equation}
 U(\{\pmb\theta_l\}; \mathbf{x}) = \left(\prod\limits_{l = 1}^L U_L(\pmb\theta_l)\right)U_E(\mathbf{x}),
 \label{eq:uni_single_enc}
\end{equation}
where $U_L(\pmb\theta_l)$ represents the trainable part of the $l$-th layer.
In many cases, we want to apply the data re-uploading technique \citep{perez2020data} to increase the performance and stability of the QNN during training. Data re-uploading means the repeated application of the encoding unitary $U_E$ at the beginning of every quantum layer, thus leading to
\begin{equation}
 U(\{\pmb\theta_l\}; \mathbf{x}) = \prod\limits_{l = 1}^L\left( U_L(\pmb\theta_l)U_E(\mathbf{x})\right).
 \label{eq:uni_dreup}
\end{equation}
Usually, we compute outcomes of multiple shots, i.e., we prepare $N$ copies of the input quantum state $\rho_{i}$ and produce $N$ copies of the output state $\rho_{f,n}$ using the same PQC defined by $U(\pmb\theta;\mathbf{x})$. To compute an action $\mathbf a$ from these, we can perform the measurement $\Pi$ on each of the $N$ final states yielding $N$ outcomes
\begin{equation}
    y_{n} \sim \Tr{\Pi\rho_{f}}.
\end{equation}
Actions can be calculated from the measurement outcomes $y_n$ with some post-measurement processing function $f_{\text{post}}$: 
\begin{equation}
    \mathbf a = f_{\text{post}}(y_1, y_2, ..., y_N).
\end{equation}
The selected actions are then executed in the classical environment leading to reward scores, and subsequently, the PPO loss $\mathcal{L}^{\text{PPO}}(\{\pmb\theta_l\})$ defined by
Eq.~\ref{eq:ppofull} is calculated on a classical machine. The PPO loss is then used to compute updated PQC parameters via a gradient descent algorithm with learning rate $\alpha$: $\pmb\theta_l \leftarrow \pmb\theta_l - \alpha \nabla_{\pmb\theta_l}\mathcal{L}^{\text{PPO}}$.

\section{Hybrid QRL in adapted latent observation spaces}
\label{sec:latent-qrl}

In this section, we introduce an end-to-end hybrid quantum-classical system and a training method for training and inference of QRL agents in classical RL environments with high dimensional observation spaces. The proposed system consists of a classical AE as a dimension reduction and feature extraction component, a QNN-based hybrid PPO agent with quantum policy and classical critic, and a classical optimizer which is jointly optimizing the parameters of both the quantum and the classical parts of the system for optimal decision making. 

\begin{figure*}[h]
    \centering
    \includegraphics[height=8cm]{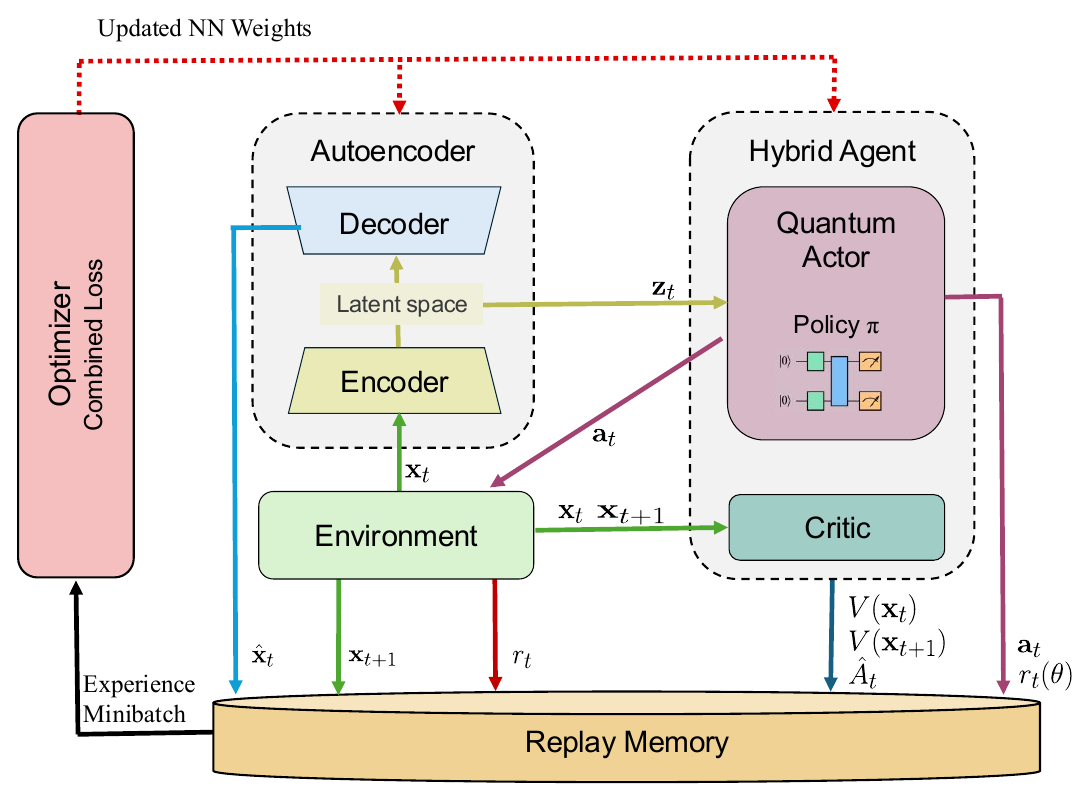}
    \caption{\textbf{Hybrid quantum-classical QRL training via the PPO algorithm in latent observation space.}
    An Autoencoder is introduced to learn latent state representation of a Classical Environment for optimal performance of a hybrid QRL PPO agent with quantum policy network.
    At each timestep, a Classical Optimizer updates the parameters for the Autoencoder and Quantum Policy networks based on a joint loss function.
    The Encoder $\mathcal{E}$ compresses the observed state $\mathbf{x}_t$ into a latent feature vector $\mathbf{z}_t$, that is passed to a Decoder $\mathcal{D}$. The reconstructed $\hat{\mathbf{x}}_t$ is used for calculating the loss function of the AE.
    The Quantum Policy $\pi$, i.e., the Actor of the QRL agent, will then choose best action $\mathbf{a}_t$ based on the latent $\mathbf{z}_t$, yielding the reward $r_t$.
    }
    \label{fig:hybrid-ae-loop}
\end{figure*}

The key component is the joint training algorithm, which ensures that the AE learns a feature extraction which is specific to the given RL problem and QPU architecture. 
The proposed architecture enables NISQ quantum agents to be used in cases where the problem's dimensionality and complexity would render simple QRL approaches infeasible.
The proposed hybrid training loop and its key components are shown on Figure~\ref{fig:hybrid-ae-loop}.
While other works explored dimension reduction methods in the context of QRL \citep{Chen_TN_AE_QRL, DimRedQRL}, a joint training of AEs and QRL agents in a hybrid quantum-classical training loop 
represents a novel solution to this problem.

\par Our method employs a PQC-based quantum agent $\pi_\theta$ defined with circuit architecture $U(\{\pmb\theta_l\};\mathbf{x})$ as explained by Eq.~\ref{eq:uni_single_enc} (or Eq.~\ref{eq:uni_dreup} with data re-uploading included), with trainable parameters $\{\pmb\theta_l\}$. Furthermore, we use an encoder network $\mathcal{E}$ with trainable parameters $\pmb\theta_{\mathcal{E}}$ and a decoder network  $\mathcal{D}$ with trainable parameters $\pmb\theta_{\mathcal{D}}$. Figure \ref{fig:hybrid-ae-loop} presents all of the components of the training loop for the proposed hybrid QRL model.

Joint training of the networks is implemented by calculating the weighted sum of PPO and AE losses with weight $c_{AE}$:
\begin{equation}
    \mathcal{L} = \mathcal{L}^\text{PPO} + c_{AE}\mathcal{L}^\text{AE}.
    \label{eq:combinedloss}
\end{equation}
The PPO loss $\mathcal{L}^\text{PPO}$ is computed as in Eq.~\ref{eq:ppofull} and $\mathcal{L}^\text{AE}$ is the AE loss suitable for the problem (e.g. MSE loss). Crucially, all trainable parameters $\{\pmb\theta_l\}$, $\pmb\theta_{\mathcal{E}}$, $\pmb\theta_{\mathcal{D}}$ are updated using the same optimizer, with gradients computed from the combined loss $\mathcal{L}$ defined by Eq.~\ref{eq:combinedloss}:
\begin{align*}
\pmb\theta_l &\leftarrow \pmb\theta_l - \alpha \nabla_{\pmb\theta_l}\mathcal{L} \nonumber \\
\pmb\theta_{\mathcal{E}} &\leftarrow \pmb\theta_{\mathcal{E}} - \alpha \nabla_{\pmb\theta_{\mathcal{E}}}\mathcal{L} \nonumber \\
\pmb\theta_{\mathcal{D}} &\leftarrow \pmb\theta_{\mathcal{D}} - \alpha \nabla_{\pmb\theta_{\mathcal{D}}}\mathcal{L}
\end{align*}

During inference, only the (trained) encoder $\mathcal{E}$ is used to compress feature vectors $\mathbf{x}_t$ into latent features $\mathbf{z}_t$, which are used by the quantum policy $\pi$ to generate actions $\mathbf{a}_t$. Optionally, it is possible to use pre-trained AEs and fine-tune them for the given control task in the proposed hybrid RL training loop. This method is called \textit{hot-starting} in contrast to starting with randomly initialized AE weights which is called \textit{cold-starting}. 

For complex feature vectors composed of different data types such as numeric, visual and categorical data, one can use separate AEs for each specific data type with their corresponding loss functions and weights contributing to the combined loss. In case of two or more AEs are used, all latent variables are encoded into the initial quantum state, with different suitable, encoding method.

\section{Experiment design for hybrid latent-space QRL}
\label{sec:numexp-setup}
We designed a series of numerical experiments to validate the proposed methods and demonstrate the advantages of the novel hybrid QRL training approach. For that purpose, we integrated both classical environment simulators for RL optimization problems and QPU backend simulators for quantum circuit evaluations into ML libraries for gradient based training methods. The autoencoders were then easily integrated into this homogeneous software design through the underlying ML framework. In reason of computation resource efficiency, we designed minimal statistical ensemble training experiments where learning curve performance analysis can already show the characteristic differences between novel and existing methods.

\subsection{Software integration}
\label{sec:simulators}
For the classical environment simulations, we used the RL benchmark environment API of the open-source Gymnasium \citep{towers_gymnasium_2023}. From the Gymnasium library we used the cart-pole problem, furthermore, we created a custom environment compatible with the Gymnasium API based on \texttt{numpy} for the visual navigation problem.

For the quantum computing simulations, we used both qubit-based and photonic architecture simulators compatible with state-of-the-art QML software frameworks, i.e., the PennyLane software framework \citep{PennyLaneBergholm} to implement qubit-based numerical experiments, and Piquasso photonic quantum simulation software platform \citep{piquasso} for continuous-variable QNN (CV-QNN) architectures \citep{KilloranCVQNN}.

Both PennyLane and Piquasso implement QPU simulator backends that are compatible with TensorFlow \citep{AbadiTF}, i.e., the experiments can be run on GPUs within HPC clusters. Furthermore, both frameworks come with out-of-the-box methods for calculating quantum gradients of PQCs via parameter-shift rule for qubit-based circuits and finite difference method for photonic circuits. Alternatively, in these experiments we implemented the exact gradients with TF from the computational graph of circuits to achieve a significant speed-up of the simulations. 

\subsection{Dimension reduction}
\label{sec:aedimred4qnns}

\begin{figure*}[!htb]
    \minipage{0.45\textwidth}%
    \centering
        \includegraphics[width=0.75\linewidth]{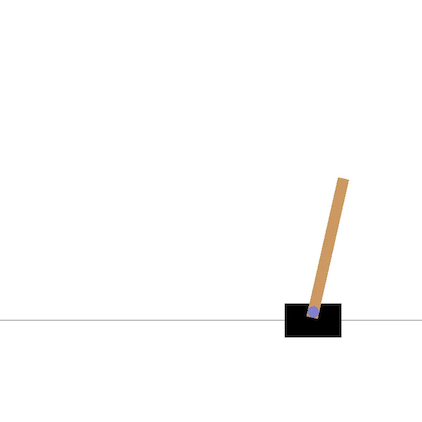}
        \caption*{(a) Cart-pole balancing problem}
    \endminipage\hfill
    \minipage{0.45\textwidth}
    \centering
        \includegraphics[width=0.75\linewidth]{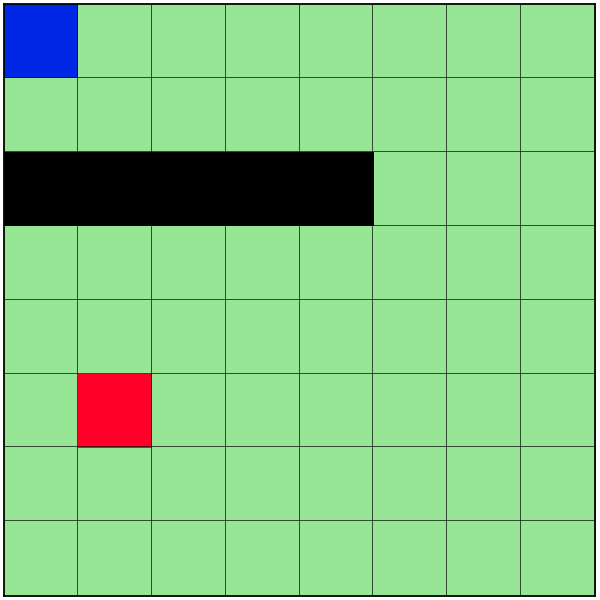}
        \caption*{(b) Visual navigation problem}
    \endminipage
    \caption{\textbf{Optimal control problems used in this work.} Two well-known classical simulated environments are selected for the numerical experiments. We used the OpenAI Gym protocol for the simulations during training of the hybrid QRL agents.
    \textbf{(a) Cart-pole balancing problem.} The goal is to keep the reverse pendulum from falling by applying left and right forces to the cart. Original observable state has 4 dimensions (cart position and velocity, pole angle and angular velocity), which we reduce to a 2-dimensional latent-space feature vector.
    \textbf{(b) Visual navigation problem.} The goal is to navigate the player (red square) to the target (blue square) while avoiding the walls (black squares). The agent receives the observable state as a $48\times48$ pixel sized grayscale image representing an $8\times8$ logical grid made from $6\times 6$ pixel sized cells.
    We reduce this original feature vector to a 8-dimensional or 4-dimensional latent vector using a convolutional AE.
    }
    \label{fig:three-environments}
\end{figure*}

We test the proposed solution on the \texttt{CartPole-v1} (Figure \ref{fig:three-environments} a.) and the \texttt{Maze-v0} (Figure \ref{fig:three-environments} b.) environments, the former is part of Gymnasium, while the latter is a custom environment. Throughout our experiments, we use fully connected AEs to compress the \texttt{CartPole-v1} observables and convolutional AEs for the \texttt{Maze-v0} environments, as these are better for visual data \citep{convAE-1, convAE-2}.
Table \ref{tab:environment-summary} summarizes these environments and their original feature space dimensionalities ($\dim(\mathbf{x})$) and the dimensions of the latent feature spaces ($\dim(\mathbf{z})$). 

\newcommand{\wc}{\cellcolor{white}}
\newcommand{\gc}{\cellcolor{gray!25}}

\begin{table*}[!htb]
  \centering
  \begin{tabular}{|>{\raggedright\arraybackslash}p{24mm}|>{\centering\arraybackslash}p{24mm}|>{\centering\arraybackslash}p{10mm}|>{\centering\arraybackslash}p{10mm}|>{\centering\arraybackslash}p{24mm}|}
    \hline
    \textbf{Environment} & \textbf{QC platform} & \textbf{dim}$(\mathbf{x})$ & \textbf{dim}$(\mathbf{z})$ & \textbf{Action space} \\
    \specialrule{1pt}{0pt}{0pt}
    \texttt{CartPole-v1} & qubit & 4 & 2 & \texttt{Discrete(2)} \\
    \hline
    \texttt{Maze-v0} & qubit & 2304 & 8 & \texttt{Discrete(4)} \\
    \specialrule{1pt}{0pt}{0pt}
    \texttt{CartPole-v1} & qumode & 4 & 2 & \texttt{Discrete(2)} \\
    \hline
    \texttt{Maze-v0} & qumode & 2304 & 6 & \texttt{Discrete(4)} \\
    \hline
  \end{tabular}
    \caption{\textbf{Simulated environments for different quantum computing platforms}. $\dim(\mathbf{x})$ 
is the original observation space dimension as defined by the environment, $\dim(\mathbf{z})$ is the
latent space dimension of the trained AE. Note that for the \texttt{Maze-v0} environment, we use $8$ qubits but only $6$ qumodes, and hence the difference in $\dim(\mathbf{z})$. Each environment has a discrete action space with different number of possible actions.}
  \label{tab:environment-summary}
\end{table*}

\subsection{QRL agent circuit designs}
The proposed latent observation space QRL method can be implemented across different quantum platforms and QPU architectures. To demonstrate this, we conduct numerical experiments using the two main approaches: qubit-based circuits and photonic (continuous-variable) quantum circuits. In this section we detail the QNN structure and measutement types for both of these architectures.

\begin{figure*}[!htb]
    \minipage{1.0\textwidth}%
       \newcommand{\GG}[2]{\gategroup[#1,steps=#2,style={dashed,rounded corners,inner xsep=2pt},background,label style={label position=below,anchor=north,yshift=-0.2cm}}
        \newcommand{\Meter}[1]{\meter{$\langle #1 \rangle$}}
        \centering
        \scalebox{.7}{\begin{quantikz}[color=black]
        \lstick{$\ket{0}$} & \gate[style={, fill=erblue!50}]{\ \ } \GG{4}{1}] {{$\begin{matrix}\textrm{\large{Angle}} \\ \textrm{\large{Embedding}}\end{matrix}$}}
                           & \qw 
                           & \gate[style={, fill=erred!50}]{{R(\alpha_1^1, \beta_1^1, \gamma_1^1)}} \GG{4}{4}] {{ ~~~~ \large First parametric layer }} 
                           & \gate[4,nwires=3,style={text width=35pt, fill=ergreen!50},label style={rotate=90}]{\hspace{-0.7cm}\text{CNOT Block 1}\hspace{-0.7cm}}
                           & \gate[style={, fill=erred!50}]{{R(\alpha_1^2, \beta_1^2, \gamma_1^2)}}
                           & \gate[4,nwires=3,style={text width=35pt, fill=ergreen!50},label style={rotate=90}]{\hspace{-0.7cm}\text{CNOT Block 2}\hspace{-0.7cm}}
                           & \qw \ldots
                           & \meter[style={fill=erorange!50}]{$Z$} \\
        \lstick{$\ket{0}$} & \gate[style={, fill=erblue!50}]{\ \ }
                           & \qw
                           & \gate[style={, fill=erred!50}]{{R(\alpha_2^1, \beta_2^1, \gamma_2^1)}}
                           & \qw
                           & \gate[style={, fill=erred!50}]{{R(\alpha_2^2, \beta_2^2, \gamma_2^2)}}
                           & \qw
                           & \qw \ldots
                           & \meter[style={fill=erorange!50}]{$Z$} \\
        \lstick{\vdots} 
                            & \vdots
                            & 
                            &
                            & 
                            & 
                            & \\
        \lstick{$\ket{0}$} & \gate[style={, fill=erblue!50}]{\ \ }
                           & \qw
                           & \gate[style={, fill=erred!50}]{{R(\alpha_N^1, \beta_N^1, \gamma_N^1)}}
                           & \qw
                           & \gate[style={, fill=erred!50}]{{R(\alpha_2^2, \beta_2^2, \gamma_2^2)}}
                           & \qw
                           & \qw \ldots
                           & \meter[style={fill=erorange!50}]{$Z$}
        \end{quantikz}}
        \vspace{0.1cm}
        \caption*{(a) Qubit-based circuit Ansatz for QRL }
    \endminipage\hfill
    \vspace{1cm}
    \minipage{1.0\textwidth}
        \newcommand{\GG}[2]{\gategroup[#1,steps=#2,style={dashed,rounded corners,inner xsep=2pt},background,label style={label position=below,anchor=north,yshift=-0.2cm}}
        \centering
        \scalebox{.7}{\begin{quantikz}[color=black]
        \lstick{$\ket{0}$} & \gate[style={, fill=erred!50}]{S(1/2)} \GG{4}{1}] {{$\begin{matrix}\textrm{\large{Initial}}\\\textrm{\large{Squeezing}}\end{matrix}$}}
                           & \qw 
                           & \qw 
                           & \gate[style={, fill=erblue!50}]{} \GG{4}{1}] {{$\begin{matrix}\textrm{\large{Displacement}}\\\textrm{\large{Encoding}}\end{matrix}$}}
                           & \qw 
                           & \gate[4,nwires=3,style={text width=35pt, fill=ergreen!50},label style={rotate=90}]{\hspace{-0.35cm}U_1(\boldsymbol{\theta}_1, \boldsymbol{\varphi}_1) \hspace{-0.35cm}} \GG{4}{5}] {{ ~~~~ \large First parametric layer }} 
                           & \gate[style={, fill=erred!50}]{{S(z_1)}} 
                           & \gate[4,nwires=3,style={text width=35pt, fill=ergreen!50},label style={rotate=90}]{\hspace{-0.35cm}U_2(\boldsymbol{\theta}_2, \boldsymbol{\varphi}_2) \hspace{-0.35cm}}
                           & \gate[style={, fill=eryellow}]{D(\alpha_1)}
                           & \gate[style={, fill=erpurple!50}]{ K(\kappa_1)}
                           & \qw \ldots
                           & \meter[style={fill=erorange!50}]{$X_\varphi$} \\
        \lstick{$\ket{0}$} & \gate[style={, fill=erred!50}]{S(1/2)}
                           & \qw
                           & \qw
                           & \gate[style={, fill=erblue!50}]{\ \ }
                           & \qw
                           & \qw 
                           & \gate[style={, fill=erred!50}]{S(z_2)}
                           & \qw 
                           & \gate[style={, fill=eryellow}]{D(\alpha_2)}
                           & \gate[style={, fill=erpurple!50}]{K(\kappa_2)}
                           & \qw \ldots
                           & \meter[style={fill=erorange!50}]{$X_\varphi$} \\
        \lstick{\vdots} 
                            & \vdots
                            & 
                            &
                            & \vdots
                            & 
                            & 
                            & \vdots
                            & 
                            & \vdots
                            & \vdots
                            & 
                            & \vdots
                            & \\
        \lstick{$\ket{0}$} & \gate[style={, fill=erred!50}]{S(1/2)}
                           & \qw
                           & \qw
                           & \gate[style={, fill=erblue!50}]{\ \ }
                           & \qw
                           & \qw 
                           & \gate[style={, fill=erred!50}]{S(z_N)}
                           & \qw
                           & \gate[style={, fill=eryellow}]{D(\alpha_N)}
                           & \gate[style={, fill=erpurple!50}]{K(\kappa_N)}
                           & \qw \ldots
                           & \meter[style={fill=erorange!50}]{$X_\varphi$}
        \end{quantikz}}
        \vspace{0.1cm}
        \caption*{(b) Photonic circuit Ansatz for QRL Actor}
    \endminipage
    \caption{
    \textbf{QNN architectures for hybrid PPO policies.} We used standard QNN architectures for the QRL Actor circuits. In both cases, the initial classical data encoding can be repeated with each layer, a technical method known as data re-uploading. \textbf{(a) Circuit Ansatz for the qubit-based policy.} Composed of an angle-embedding layer at the beginning of the circuit, followed by a number of standard strongly entangling trainable layers.
    \textbf{(b) Circuit Ansatz for the photonic policy.} Composed of an initialization by squeezing gates and displacement encoding layer followed by a number of trainable standard CV-QNN layers.
    }
    \label{fig:quantum-circuits}
\end{figure*}

\subsubsection*{Qubit-based PPO agent design}\label{sec:qcirc-arch}
For our qubit-based experiments, we use a QNN policy architecture based on strongly entangling layers as described in \citep{SchuldStronglyEntangling}. Such layers consist of a set of trainable rotation gates $R(\alpha, \beta,\gamma)$ applied to every qubit of the system, followed by a CNOT block to entangle qubits, then another set of rotation gates and finally a second CNOT block with different structure from the first one. These strongly entangling layers are repeated multiple times. 
To encode latent features, we use angle embedding, i.e., the encoding unitary is 
\begin{equation}
    U_E(\mathbf{z}) = \bigotimes\limits_{j=1}^M R_Y^{(j)}(f_{\text{prep}}(z_j)),
\end{equation}
where $R_Y^{(j)}$ denotes an $Y$-rotation applied to the $j$-th qubit of the system, and $f_{\text{prep}}$ is some preprocessing function. Initially all qubits are in $\ket{0}$ state, and the angle-embedding $U_E$ can be repeated at the beginning of every trainable layer as in \citep{perez2020data}. Crucially, the maximum dimension of latent feature vectors $\mathbf{z}$ is limited by the number of qubits $M$. Figure \ref{fig:quantum-circuits} (a) shows the qubit based QNN architecture without data re-uploading.

To construct actions from the output quantum state, we estimate the expectation values $\langle Z_j \rangle$ for each qubit, and then construct the discrete probability distribution 
\begin{equation}
    \Pr(a=j) = \frac{\exp\langle Z_j\rangle}{\sum\limits_k \exp  \langle Z_k\rangle},
\end{equation}
from which we sample actions $a_t$ at each timestep.

\subsubsection*{Photonic PPO agent design}\label{sec:qcirc-arch}
Besides qubit-based QNNs, we also test our method with quantum policies implemented as continuous variable QNNs (CV-QNNs), similarly to the numerical experiments presented in \citep{dnagy}. The CV-QNN Ansatz is composed of the following photonic gates: Gaussian gates including squeezing ($S(r, \phi)$), displacement $D(\alpha, \phi)$, and rotation ($R(\phi)$) a.k.a.~phase-shifting; furthermore, in order to introduce non-linearity we need at least one nonlinear CV quantum gate \citep{KilloranCVQNN}. We use Kerr-nonlinearities for this purpose ($K(\kappa)$).
The general CV-QNN circuit architecture is shown on Figure~\ref{fig:quantum-circuits} (b).
In all of the CV-QNN-based experiments, we use a pre-squeezed initial state i.e., we apply a squeezing of magnitude $1/2$ and phase $\pi/2$ to the $M$-mode vacuum state as initialization:
\begin{align}
    \ket{\psi_i} = S\left(\frac{1}{2}, \frac{\pi}{2}\right)^{\otimes M} \ket{0^{\otimes M}}.
\end{align}
Subsequently, we use displacement-encoding to encode latent features into the quantum state. Displacement-encoding applies a displacement of magnitude $f_{\text{prep}}(z_j)$ and phase $\pi/2$ to the $j$-th mode of the CV quantum system:
\begin{equation}
    U_E(\mathbf{z}) = \bigotimes_{j=1}^{M} D_j\left(f_{\text{prep}}(z_j),\frac{\pi}{2}\right),
\end{equation}
where $f_{\text{prep}}$ is some feature preprocessing function.
$f_{\text{prep}}$ can be the identity function, or an $\arctan$-based preprocessing like in \citep{dnagy}:
\begin{equation}
    f_{\text{prep}}(x) =  \begin{cases}
    + \frac{4}{\pi} |\arctan(x)|^{1/3}&~x > 0,\\
    - \frac{4}{\pi} |\arctan(x)|^{1/3}&~x < 0,\\
    0&~x=0,
    \end{cases}
    \label{eq:feature_preproc}
\end{equation}
 which ensures that the values of the displacement gate parameters remain reasonably small.

As in Eq.~\ref{eq:uni_dreup}, the encoding unitary $U_E$ can be repeated at the beginning of each trainable layer. The maximum number of modes $M$ we can simulate limits the dimension of latent features $\mathbf{z}$ that can be encoded into the system via displacement-encoding. A CV-QNN Ansatz without data re-uploading is shown on Figure \ref{fig:quantum-circuits} (b).

To construct actions from the output quantum state, we perform homodyne measurements and estimate the expectation values $\langle P_j \rangle$ for each qumode, and then construct the discrete probability distribution 
\begin{equation}
    \Pr(a=j) = \frac{\exp\langle P_j\rangle}{\sum\limits_k \exp  \langle P_k\rangle},
\end{equation}
from which we sample actions $a_t$ at each timestep.

\section{Evaluation of learning efficiency and performance}
\label{sec:numexp-results}

To enable performance comparison of training methods across different environments, we introduce a normalized version of the well-known area under learning curve (AULC) metric.
First, we consider a normalized reward by the $P\%$ of the optimal score of the given environment.
Second, we further normalize over the time dimension as well, i.e., the episode count, as follows: we consider the episode number $e_P^E$ at which the best agent of the ensemble training for a given environment $E$ reaches the previously set $P$ percent of the optimal reward score. Hence the best agent will reach the maximal normalized reward of one at normalized episode number of one.

In this work, we apply this normalization technique over the averaged learning curves of an ensemble of agents, each also smoothed using a moving window average to counteract fluctuations.

Note that for complex learning curves, the AULC and normalized AULC scores could be misleading in itself, as it is reducing performance characteristics into a single number. Hence these results always need to be assessed together with the learning curves. Fast learning leads to a convex learning curve, i.e., a normalized AULC larger than $0.5$ while value under $0.5$ could still reach $P$ percent but with a slower initial learning and a concave learning curve.

Due to the stochastic nature of RL, it is required to average learning curves over multiple agents with the same hyper-parameters but initialized with different random seeds. We train $8$ agents in parallel for each experiment, and we calculate performance metrics for $5$ out of $8$, as the outlier agents were discarded (usually without a visible learning trend).

\subsection{Demonstration of the joint training method performance}

\begin{figure*}[!htb]
\centering
\includegraphics[width=0.95\textwidth]{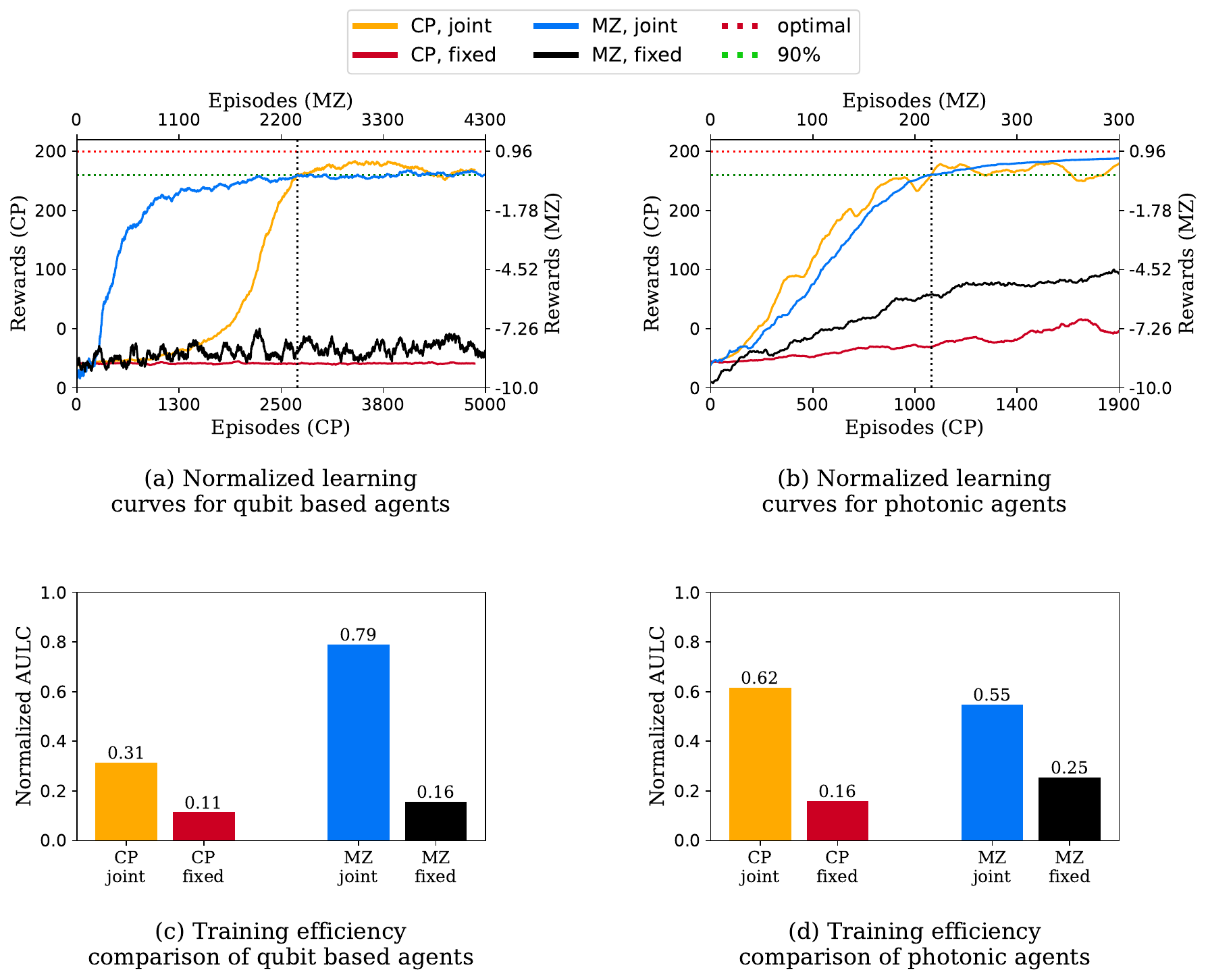}
\caption{
\textbf{Performance evaluation of the novel latent-space QRL training method.} We compare the novel latent-space training method via jointly trained AE for quantum policy against using a fixed AE for simple feature extraction. The results are shown for two types of QRL agent trained across two control problems: qubit-based and photonic agents in for the Cart-Pole balancing problem (\texttt{CartPole-v1} environment, labeled CP) and the visual navigation problem (\texttt{Maze-v0} environment, labeled MZ). \textbf{(a)-(b) Normalized learning curves.} Results show rewards normalized by each environment's optimum. Learning curves are averaged over $8$ agents, smoothed with $100$-episode moving window. \textbf{(c)-(d) Training efficiency comparisons.} Results present the normalized area under learning curve (AULC) until faster training method reaches $90\%$ of the normalized optimum. Higher values indicate faster convergence by episode normalization per environment, but we keept separate comparison per quantum agent type.}
 \label{fig:tech_compare_auc_combined}
\end{figure*}

First, we demonstrate a small-scale implementation of latent QRL training. We investigate whether the joint training of an AE and a QRL agent yields satisfactory agent performance results, and compare these with training a QRL agent using a pre-trained but fixed AE. 

These tests were executed using both qubit-based and photonic QNN policies and on both the \texttt{CartPole-v1} and the \texttt{Maze-v0} environments. The learning ability of the joint training method is further evaluated by computing the normalized AULC metric for each training session, based on the $90\%$ score thresholds. 

Resulting learning curves are presented on Figures~\ref{fig:tech_compare_auc_combined} (a) and (b) for qubit-based and photonic architectures respectively. Furthermore the computed normalized AULC scores are shown on Figures~\ref{fig:tech_compare_auc_combined} (c) and (d) for qubit-based and photonic architectures respectively. Both the normalized learning curves and AULC socres indicate the necessity of joint training to achieve satisfactory learning performance: agents achieve at least 90$\%$ of the optimal score only when the joint training is applied. These results collectively show that the implementation of the proposed hybrid QRL agent with joint training is able to learn and solve RL problems even with high-dimensional observation spaces like \texttt{Maze-v0}.

Furthermore, results indicate that compared to the fixed pre-trained AE solution, the joint training approach shows better learning ability, and in many cases, it is able to solve RL problems that a fixed AE approach is unable to solve. A possible explanation for this phenomenon is that when starting the process with a pre-trained AE, it is close to a local minimum (but possibly far from a global one), and therefore the AE and the agent struggle to get out from this local minimum. However, confirming this hypothesis would require further numerical experimentats, beyond the scope of this paper.

Detailed description of the experiments and hyperparameters are summarized in Appendix~\ref{appendix:suppl}.


\subsection{Exploration of AE compressive power vs QNN layer count}

\begin{figure*}[!htb]
\centering
\includegraphics[width=0.9\textwidth]{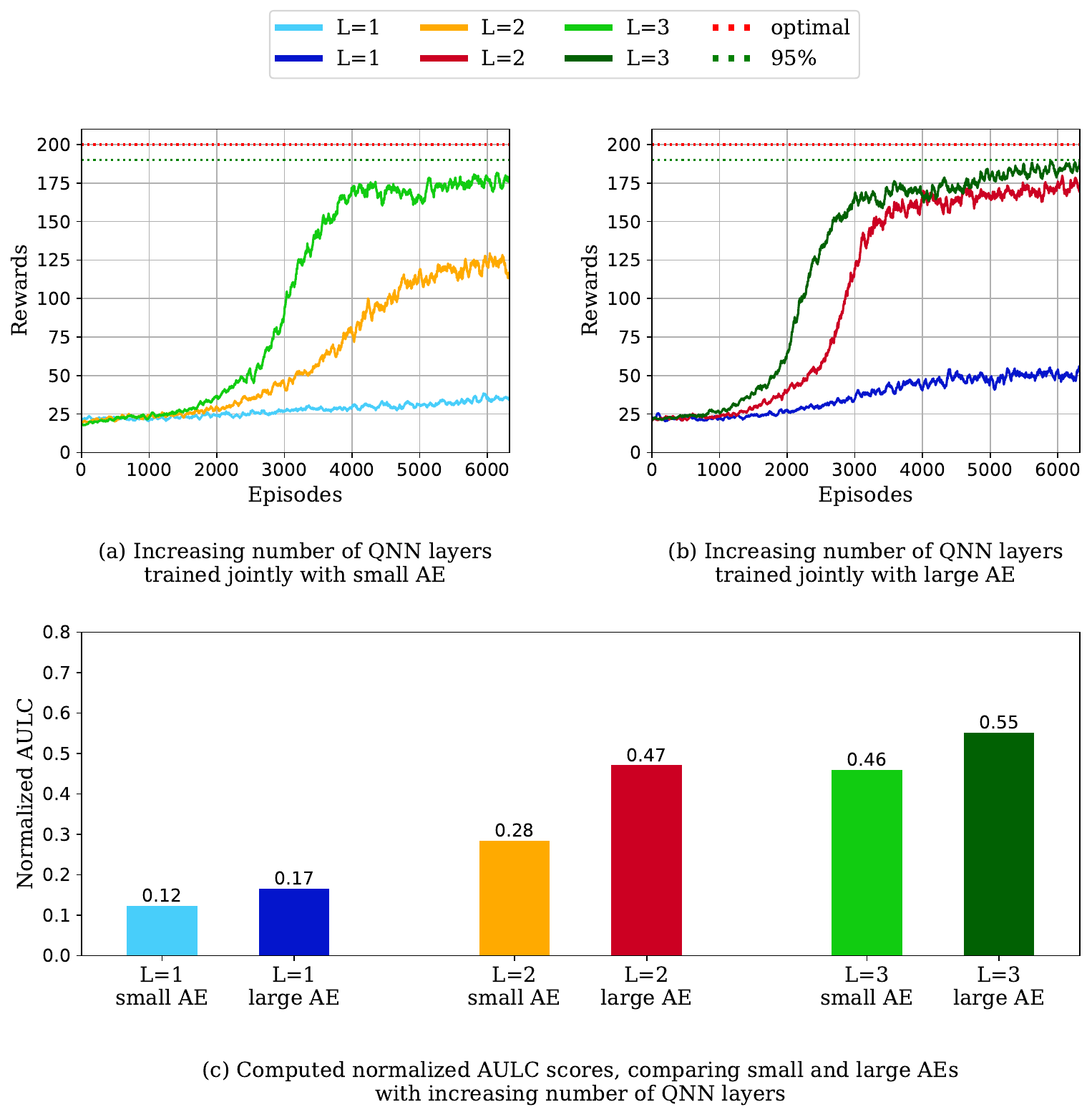}
\caption{\textbf{Learning power of hybrid agents: interdependence of QNN and AE complexity.} Numerical experiments show that by increasing the number of AE parameters smaller QNN agents are enabled to reach better performance. Similarly, with increasing number of QNN layers, a smaller AE is sufficient for convergence. 
Results show that the minimal single-layer QNN agent is unable to learn even with large AE, however when we consider a two-layer QNN, using the larger AE enables satisfactory learning performance. For large QNNs, increasing the AE size does not yield significantly better results. Simulations were performed on the \texttt{CartPole-v1} environment using qubit-based agents.
}
 \label{fig:ae_pcount_vs_qnn_lcount_combined}
\end{figure*}

Besides testing the joint training method against a fixed AE solution, we also explore the interdependence between the size of the AE and the number of QNN layers.
Balancing AE and quantum resources might be necessary because of compute resource limitations, or to maximize the utility of both quantum and classical resources. 

Specifically, we use the \texttt{CartPole-v1} environment to assess how variations in the size of the AE and the number of QNN layers affect learning performance. Due to the computational cost of simulating photonic QNNs, this study is done using qubit-based policies only.

To explore the interdependence between AE and QNN complexities, we simulate a set of agents with various configurations as follows. We run a number of training experiments in which we vary the number of QNN layers, with a fixed AE size. We repeat these simulations with a small AE and a larger AE configuration.

Learning curves for the small AE simulation are shown on Figure \ref{fig:ae_pcount_vs_qnn_lcount_combined} (a) while Figure \ref{fig:ae_pcount_vs_qnn_lcount_combined} (b) shows these in the case of the larger AE. 
Furthermore, the computed the normalized AULC values in each case are shown on Figure \ref{fig:ae_pcount_vs_qnn_lcount_combined} (c).

Results of these numerical experiments support our hypothesis that as the size of the AE increases, smaller QNN configurations yield improved performance, and similarly, as we increase the number of QNN layers, a smaller AE is enough to achieve satisfactory learning performance. However due to the lack of computational resources, these experiments were only conducted on the \texttt{CartPole-v1} environment with qubit-based agents, thus finding whether a similar interdependence  also exists in the case of the \texttt{Maze-v0} environment or in the case of photonic agents, further numerical experiments are required.

Thes results show that increasing the AE allows for a reduction in the number of QNN layers required, as expected. Furthermore, we find the limit of under-dimensioned QNN will not learn to solve the problem, not even with large AEs.

However, we find that there are limits to this trade-off: for instance, while a larger AE can compensate for fewer QNN layers to some extent, it is insufficient for extremely small QNN architectures such as single-layer QNNs. On the other hand, with an overdimensioned AE, independent of the QNN depth, the agent will remain classically determined, without performance explained by the QNN's representative power. With balanced AE sizing, in the case of the \texttt{CartPole-v1} problem, we find that a minimum of three QNN layers is necessary to achieve satisfying learning results. Detailed description of the experiments and hyperparameters are summarized in Appendix~\ref{appendix:suppl}.

\subsection{Benchmark results against classical baselines}
\begin{figure*}[!htb]
    \centering
\includegraphics[width=0.9\textwidth]{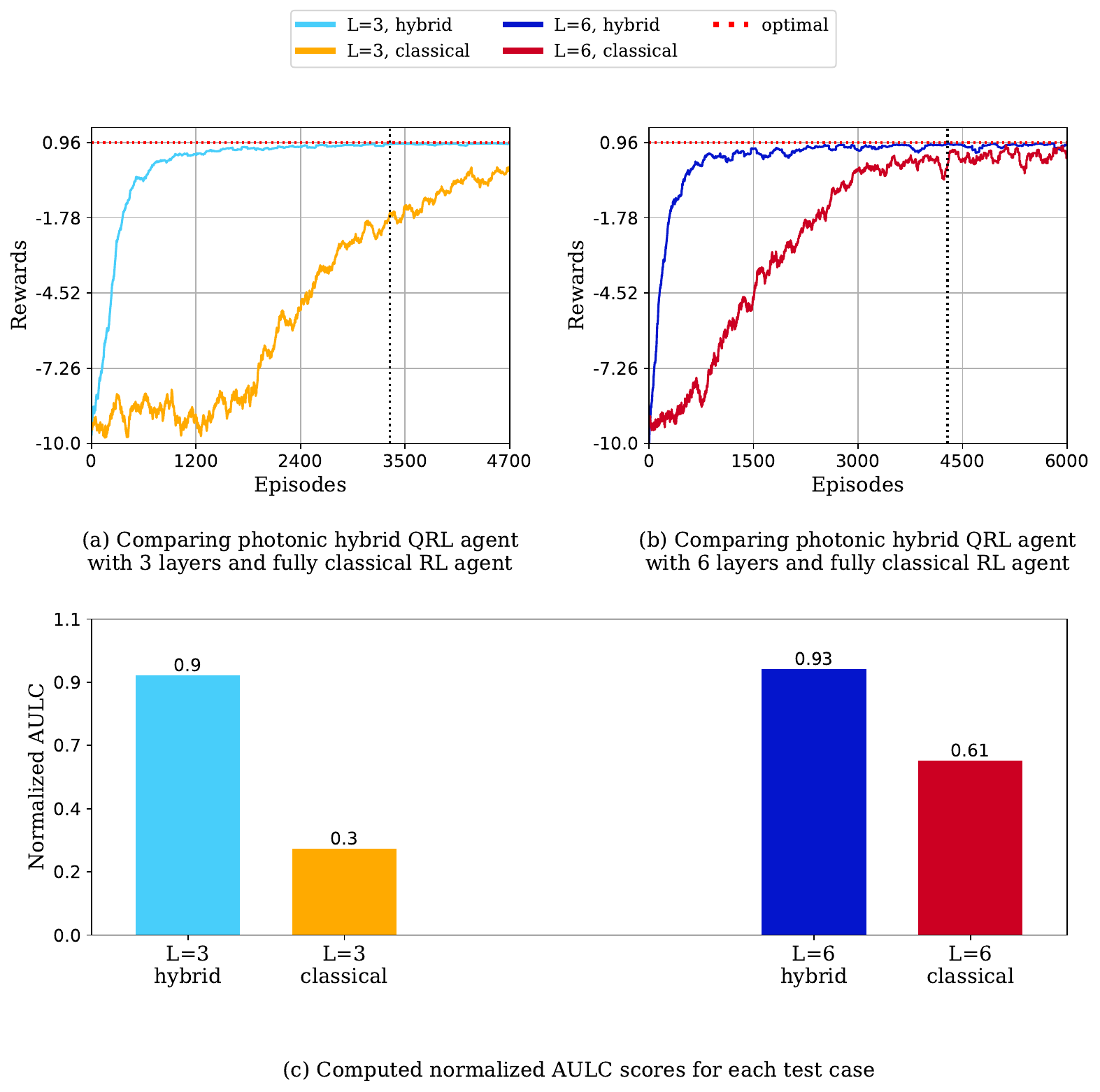}
    \caption{
    \textbf{Comparison of fully classical and hybrid agents.} Comparing learning curves for agents with similar parameter counts and hyperparameters and randomly initialized parameters. Figures show the smoothed learning curves averaged over 8 parallel agents.  \textbf{(a) - (b)}: results for photonic QNN agents; \textbf{(c)}: Computed AULC scores for each case. For shallow photonic QNN-based agents, the difference between the hybrid solution and its classical counterpart seems larger than for deeper circuits.
    }
    \label{fig:maze-classic-vs-q-layerwise}
\end{figure*}

To compare our solution with a fully classical counterpart, we choose the \texttt{Maze-v0} environment, as this is only solvable by a quantum agent when an AE is assisting with feature extraction due to the high dimensionality of the problem. 
Furthermore, this study uses photonic agents, as qubit-based agents with the  shallow Ansatz configuration did not achieve near-optimal performance.

To compare the classical and hybrid solutions, the number of trainable parameters in the classical agent is adjusted so that the combined number of parameters in the QNN and the AE of the hybrid agent is approximately the same. Trainable parameter counts for each scenario are detailed in Appendix~\ref{appendix:suppl}.

Figures \ref{fig:maze-classic-vs-q-layerwise} (a) and (b) present the comparison of learning curves between a photonic quantum agent assisted by a convolutional AE and a fully classical CNN-based agent. Configurations presented on Figures \ref{fig:maze-classic-vs-q-layerwise} (a) and (b) differ in the number of QNN layers used ($L=3$ and $L=6$ respectively), and the CNN parameter count is adjusted to the number of QNN params in each case.

Results show that the photonic agents acheive faster learning compared to a similarly sized fully classical agent. This is also confirmed by the computed normalized AULC scores shown on Figure~\ref{fig:maze-classic-vs-q-layerwise}~(c). Detailed description of the experiments and hyperparameters are summarized in Appendix~\ref{appendix:suppl}.

\subsection{Scalability and noise resistance}
\begin{figure*}[!htb]
    \minipage{0.475\textwidth}%
    \centering
        \includegraphics[width=0.99\linewidth]{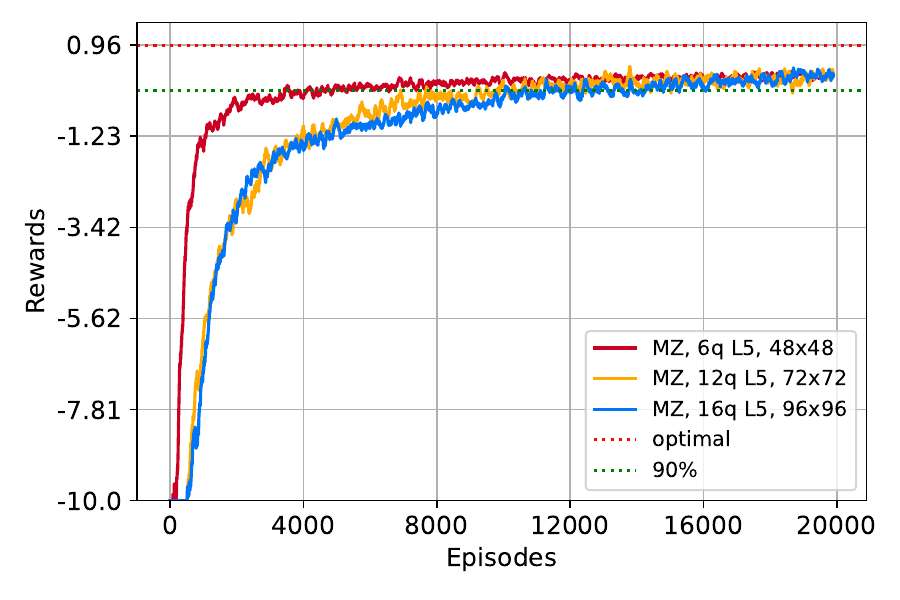}
        \caption*{\textbf{(a)}}
    \endminipage\hfill
    \minipage{0.475\textwidth}
    \centering
        \includegraphics[width=0.99\linewidth]{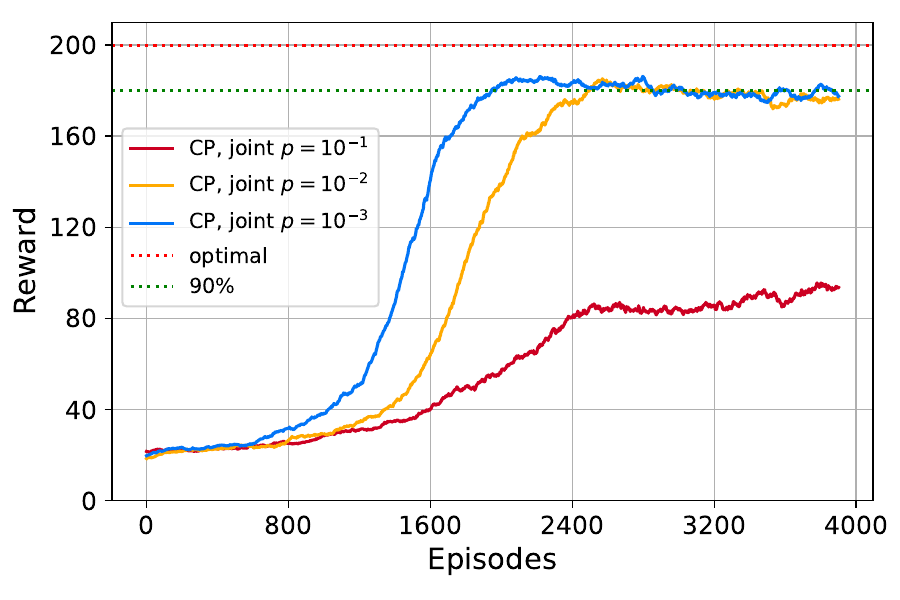}
        \caption*{\textbf{(b)}}
    \endminipage
    \caption{Scalability test results (a): learning curves for the \texttt{Maze-v0} environment when scaling from 6 to 12 and 16 qubits with proportionally larger input images (48$\times$48 to 72$\times$72 and 96$\times$96). Noise-resilience test (b): learning curves for the \texttt{CartPole-v1} environment under different depolarizing noise probabilities.}
    \label{fig:noise-and-scaling}
\end{figure*}

We performed additional simulations to evaluate both the scalability of our method and its robustness against noise. First, we extended the \texttt{Maze-v0} experiments to larger quantum latent spaces of 12 and 16 qubits, up from the original 6 qubits. Concurrently, we scaled the problem size from $48 \times 48$ to $72 \times 72$ and $96 \times 96$ pixels for the 12- and 16-qubit simulations, respectively. As shown in Figure~\ref{fig:noise-and-scaling}~(a), the learning curves remain consistent across these larger settings, illustrating the ability of the method to scale with increasing input dimensionality and quantum model size.

Second, we investigated noise resilience by re-running the \texttt{CartPole-v1} task on a noisy qubit simulator. A depolarizing channel, characterized by the Kraus operators 
\begin{align*}
K_0 &= \sqrt{1-p} \begin{bmatrix}
        1 & 0 \\
        0 & 1
        \end{bmatrix}, \quad
K_1 = \sqrt{\frac{p}{3}}\begin{bmatrix}
        0 & 1 \\
        1 & 0
        \end{bmatrix},\\
K_2 &= \sqrt{\frac{p}{3}}\begin{bmatrix}
        0 & -i \\
        i & 0
        \end{bmatrix}, \quad
K_3 = \sqrt{\frac{p}{3}}\begin{bmatrix}
        1 & 0 \\
        0 & -1
        \end{bmatrix},
\end{align*}
was introduced prior to each strongly entangling layer of the QNN. We tested a range of depolarization probabilities ($p=10^{-1}$, $5\times10^{-2}$, $3\times10^{-2}$, $10^{-2}$, $10^{-5}$) and observed that higher error rates degraded learning performance, while lower error rates yielded curves similar to the noiseless setting. Figure~\ref{fig:noise-and-scaling}~(b) illustrates these effects.
Details of the noise-resilience and scalability tests are summarized in Appendix~\ref{appendix:suppl}. 

\section{Conclusions and outlook}\label{sec:conclusion}

Quantum machine learning offers a way of using the powerful expressivity of quantum systems for classical problem solving. However, due to the scarcity of quantum resources, an essential challenge is to efficiently encode the high-dimensional descriptors of real-world problems for quantum agents.
A natural way of addressing this challenge is to extract the task-specific information from the general descriptors.  Since such problem representations are among the most fundamental building blocks on the path towards useful QML, this work focused on a novel solution for efficiently encoding high-dimensional observations into the initial quantum state of the QRL policy circuit Ansatz for optimal control performance. 

In particular, we proposed integrating classical AEs with QRL agents in a joint hybrid quantum-classical training loop. This joint training ensures that the AE learns feature extraction tailored to the specific RL problem and QPU architecture, thereby leading to more efficient and effective learning.

 We designed a series of experiments for the comparative analysis of the proposed method for two classical RL benchmark  problems, the cart-pole balancing and a visual navigation problems. Using the particular \texttt{CartPole-v1} and \texttt{Maze-v0} environments, we implemented hybrid PPO agents in both qubit-based and photonic quantum circuits. The analysis showed that the joint training approach significantly outperforms the fixed pre-trained AE case in these two very distinct environments for qubit-based as well as photonic agents. In presenting the design of the numerical experiments, we also highlighted the importance of the interplay between the compressive power of the AE component and the control performance of the QNN agent - in relation to the specifics of each control environment.
 
 Comparisons against fully classical agents in the \texttt{Maze-v0} environment revealed that our approach of joint training reaches similar solution quality with often a faster convergence. This effect was particularly pronounced in the photonic QNN experiments. Throughout these comparative experiments, we trained agents with similar number of trainable parameters and similar hyperparameters.
 
 We tested the scalability of our solution via simulations to show that the proposed hybrid quantum RL framework scales effectively to larger latent spaces and corresponding increases in problem size. By extending from 6 to 12 and 16 qubits, the agent maintains consistent learning curves, demonstrating that the method remains robust at higher dimensionalities.
 
 Furthermore, we also tested noise resilience of the approach by introducing a depolarizing channel into the qubit-based QNN circuits. While higher noise levels degrade performance, lower noise settings yield behavior similar to the noiseless case. These results indicate that our approach can tolerate moderate levels of noise, reinforcing its potential for near-term quantum devices.

 The presented novel hybrid training method paves the way towards real-world problems where one could benefit from the expressive power of quantum agents. This initial study could be extended in multiple ways. We can generalize the study by replacing simulators with QPU prototype hardware or testing the method in real enviroments. We can also extend the method towards more complex problem spaces, e.g., multimodal observables or multi-agent scenarios, where each modality or agent will be trained with their own respective AEs. It would be interesting to see whether the observed faster convergence would also prevail in these situations.
 
 Our results show that joint training performs better with a randomly initialized AE than with a pre-trained AE. This raises an open question: does pre-training introduce biases or restrict adaptability in joint optimization? One possibility is that a randomly initialized AE allows for better co-adaptation, while a pre-trained AE may overfit to specific features. Further investigation is needed to understand this effect and optimize the integration of pre-training and joint training.
 
\section*{Acknowledgment}
D\'aniel T. R. Nagy, Bence Bak\'o, Zolt{\'a}n Zimbor{\'a}s and Zs{\'o}fia Kallus would like to thank the support provided by the Ministry of Culture and Innovation of Hungary from the National Research, Development and Innovation Fund (NKFIH), financed under the KDP-2021 and KDP-2023 funding schemes (Grants No. C1788111 and C2245275). Zolt{\'a}n Zimbor{\'a}s would like to thank the support of the NKFIH through  the Grants No. FK 135220 and TKP2021-NVA-29 and the Quantum Information National Laboratory of Hungary. The authors acknowledge the computational resources provided by the Wigner Scientific Computational Laboratory.

\newpage
\appendix
\section{Supplementary Information}
\label{appendix:suppl}
In this section, we present supplementary information detailing the configurations utilized in our experiments with the \texttt{CartPole-v1} and \texttt{Maze-v0} environments.
The tables provided here detail the QNN parameter counts for both qubit and qumode platforms as well as of the parameter counts for the classical fully connected and convolutional AEs. Furthermore, we detail the configurations and parameter counts of the fully classical CNN-based agents we use as baseline for the classical versus hybrid QRL comparisons.

Table \ref{tab:cp-policy-params} and Table \ref{tab:maze-policy-params} summarize the parameter counts for QNNs with different number of layers, distinguishing between the qubit-based and photonic agent approaches in both environments. 

\begin{table*}[!htb]
  \centering
  \begin{tabular}{|>{\raggedright\arraybackslash}p{24mm}|>{\centering\arraybackslash}p{14mm}|>{\centering\arraybackslash}p{20mm}|>{\centering\arraybackslash}p{24mm}|}
    \hline
    \textbf{Environment} & \textbf{platform} & \textbf{QNN layers} & \textbf{QNN params} \\
    \specialrule{1pt}{0pt}{0pt}
    \texttt{CartPole-v1} & qubit & 1 & 6\\
    \hline
    \texttt{CartPole-v1} & qubit & 2 & 12\\
    \hline
    \texttt{CartPole-v1} & qubit & 3 & 18\\
    \hline
    \texttt{CartPole-v1} & qubit & 6 & 32\\
    \specialrule{1pt}{0pt}{0pt}
    \texttt{CartPole-v1} & qumode & 1 & 14\\
    \hline
    \texttt{CartPole-v1} & qumode & 3 & 42\\
    \hline
    \texttt{CartPole-v1} & qumode & 6 & 84\\
    \hline
  \end{tabular}
    \caption{\textbf{Parameter counts used in the \texttt{CartPole-v1} problem.} The table summarizes the number of layers and corresponding number of QNN parameters for both qubit-bases agents and photonic agents.}
  \label{tab:cp-policy-params}
\end{table*}

\begin{table*}[!htb]
  \centering
  \begin{tabular}{|>{\raggedright\arraybackslash}p{24mm}|>{\centering\arraybackslash}p{14mm}|>{\centering\arraybackslash}p{20mm}|>{\centering\arraybackslash}p{24mm}|}
    \hline
    \textbf{Environment} & \textbf{platform} & \textbf{QNN layers} & \textbf{QNN params} \\
    \specialrule{1pt}{0pt}{0pt}
    \texttt{Maze-v0} & qubit & 1 & 24\\
    \hline
    \texttt{Maze-v0} & qubit & 3 & 72\\
    \hline
    \texttt{Maze-v0} & qubit & 5 & 120\\
    \specialrule{1pt}{0pt}{0pt}
    \texttt{Maze-v0} & qumode & 1 & 94\\
    \hline
    \texttt{Maze-v0} & qumode & 3 & 282\\
    \hline
    \texttt{Maze-v0} & qumode & 6 & 564\\
    \hline
  \end{tabular}
     \caption{\textbf{Parameter counts used in the \texttt{Maze-v0} problem.} The table summarizes the number of layers and corresponding number of QNN parameters for both qubit-bases agents and photonic agents.}
  \label{tab:maze-policy-params}
\end{table*}

To investigate the interdependence of QNN layer count and AE param counts, we trained the hybrid QRL agent together with a small and a large AE using 1, 2 and 3 QNN layers. Table \ref{tab:cp-dense-ae-params} contains the parameter counts for the small and large fully-connected AEs. The small AE consists of a single \texttt{Dense(2)} layer followed by a \texttt{sigmoid} activation, while the large AE consists of a \texttt{Dense(8)} and a \texttt{Dense(2)} layer, each followed by \texttt{sigmoid} activations.
Both AEs are pretrained using a learning rate schedule defined by the \texttt{PiecewiseConstantDecay} class. The decay boundaries are set at 250, 500, 750, 1000, 1250, and 1500 epochs, with corresponding learning rate values of 0.05, 0.01, 0.005, 0.001, 0.0005, 0.0001, and 0.00005, respectively. Both AEs are trained for 20,000 epochs with a batch size of 16.

\begin{table*}[!h]
    \centering
    \begin{tabular}{|c|c|c|c|c|}
        \hline
        \textbf{Environment} & \textbf{Name} & \textbf{Hidden Sizes} & \textbf{Encoder Params} & \textbf{Decoder Params} \\
        \specialrule{1pt}{0pt}{0pt}
        \texttt{CartPole-v1} & small AE & [2] &  10 &  12\\ 
        \hline
        \texttt{CartPole-v1} & large AE & [8,2] & 58 & 60 \\ 
        \hline
    \end{tabular}
    \caption{\textbf{Parameter counts for the fully connected AEs used in the \texttt{CartPole-v1} experiments}. The table summarizes the hidden layer configuration, encoder and decoder parameter counts for the small and large fully-connected AEs used in the \texttt{CartPole-v1} experiments. 
    }
    \label{tab:cp-dense-ae-params}
\end{table*}

\begin{table*}[!h]
    \centering
    \begin{tabular}{|c|c|c|c|c|c|}
        \hline
        \textbf{Name} & \textbf{Platform} & \textbf{Filter Sizes} & \textbf{Pooling size} & \textbf{Encoder Params} & \textbf{Decoder Params} \\
        \specialrule{1pt}{0pt}{0pt}
        convAE & qumode & [2, 2] & 4 & 172 & 221 \\ 
        \hline
        convAE & qubit & [2, 2] & 4 & 210 & 257 \\ 
        \hline
    \end{tabular}
    \caption{\textbf{Parameter counts for the convolutional AE used in the \texttt{Maze-v0} experiments.} The table summarizes the number of parameters and filter configurations of the convolutional AE for both qubit based agents and photonic agents.}
    \label{tab:mz-conv-ae-params}
\end{table*}

Table \ref{tab:mz-conv-ae-params} details the hyperparameters of the convolutional AE used for the \texttt{Maze-v0} environment. This consists of two primary components: a convolutional encoder and a convolutional decoder. The encoder accepts input images with dimensions of 48x48 pixels and 1 channel, and encodes them into a latent space of 6 dimensions. The encoder employs \texttt{Conv2D} layers with filters set to [2, 4, 8, 8] and uses the \texttt{ReLU} activation function. There are no additional hidden layers in the encoder.

The decoder mirrors the encoder's structure. It reconstructs the 48x48 pixel output images with 1 channel from the latent space representation. The decoder uses \texttt{Conv2DTranspose} filters set to [8, 8, 4, 2] and applies bilinear interpolation for upsampling. It also uses the \texttt{ReLU} activation function and includes no extra hidden layers.

The convolutional AE is pretrained using the \texttt{PiecewiseConstantDecay} learning rate schedule, with boundaries at 250, 500, 750, 1000, 1250, and 1500 epochs. The corresponding learning rate values are 0.05, 0.01, 0.005, 0.001, 0.0005, 0.0001, and 0.00005, respectively. The convolutional AE is trained for 20,000 epochs with a batch size of 32.

\begin{table*}[!h]
    \centering
    \begin{tabular}{|c|c|c|c|c|c|}
        \hline
        \textbf{Platform} & \textbf{Number of layers} & \textbf{CNN param count} & \textbf{convAE + QNN param count} \\
        \specialrule{1pt}{0pt}{0pt}
        qumode & 1 & 519 & 487 (94 + 172 + 221) \\
        \hline
        qumode & 3 & 731 & 675 (282 + 172 + 221) \\
        \hline
        qumode & 6 & 974 & 957 (564 + 172 + 221) \\
        \specialrule{1pt}{0pt}{0pt}
        qubit & 1 & 517 & 491 (24 + 210 + 257) \\
        \hline
        qubit & 3 & 556 & 539 (72 + 210 + 257) \\
        \hline
        qubit & 5 & 609 & 587 (120 + 210 + 257) \\
        \hline
    \end{tabular}
    \caption{\textbf{Parameter counts used in the classical vs quantum benchmark experiments on the \texttt{Maze-v0} environment.} The table summarizes parameter counts for the fully classical CNN agents, and parameter counts for the hybrid quantum-classical agents including the QNN and the classical convolutional AE.}
    \label{tab:mz-cnn-vs-hybrid-params}
\end{table*}
\begin{table*}[!h]
    \centering
    \begin{tabular}{|c|c|c|c|c|c|}
        \hline
         \textbf{Environment} & \textbf{qubits} & \textbf{layers} & \textbf{QNN params} & \textbf{Encoder params} & \textbf{Decoder params} \\
        \specialrule{1pt}{0pt}{0pt}
        Maze $48\times 48$ & 6 & 5 & 90 & 1414 & 2057 \\
        \hline
        Maze $72\times 72$ & 12 & 5 & 180 & 454 & 511 \\
        \hline
         Maze $96\times 96$ & 16 & 5 & 240 & 1226 & 1319 \\
        \hline
    \end{tabular}
    \caption{\textbf{Details of scaling test experiments.} Hyperparameters and trainable parameter counts for the experiments we ran to test scalability of our approach on the \texttt{Maze-v0} environment (Figure~\ref{fig:noise-and-scaling} a).}
    \label{tab:mz-qubit-scaling}
\end{table*}

The fully classical CNN agent has input observation space consisting of images with dimensions 48x48 pixels and 1 channel and outputs a 4-dimensional vector. The CNN consists of three convolutional layers, with filter sizes that vary depending on the specific configuration. The number of channels in the \texttt{Conv2D} layers are 8, 4, and 3, respectively. After the \texttt{Conv2D} layers, the output is flattened, followed by a single hidden dense layer whose size also varies based on the chosen configuration. The agent uses the \texttt{ReLU} activation function throughout the network.
Table \ref{tab:mz-cnn-vs-hybrid-params} compares parameter counts of fully classical CNN agents against parameter counts of hybrid quantum-classical agents, we use in the classical versus hybrid QRL agent benchmark tests.

\subsection*{Noise-resilience tests}
For the noise-resilience tests, we used a QNN architecture very similar to the noise-free case, shown on Figure~\ref{fig:quantum-circuits}(a) with an additional depolarizing channel on every qubit after every parametric layer implemented in Pennylane by the \texttt{qml.DepolarizingChannel} class. We use three consecutive layers with a total of $18$ trainable quantum parameters and we use depolarizing probabilities $p\in\{10^{-1}, 10^{-2}, 10^{-3}\}$. The classical encoder and decoder networks had $114$ and $116$ trainable parameters respectively.

\subsection*{Hardware usage and code}
Simulations were run on a hardware equipped with an Intel(R) Xeon(R) CPU E5-2650 CPU clocked at 2.00 GHz, 32GB memory and an NVIDIA GeForce GTX 1080 Ti GPU with 12GB VRAM. The scripts, configuration files, and resulting data files used in this study are available at:  \url{https://github.com/Budapest-Quantum-Computing-Group/hybrid_latent_qrl}.

\bibliographystyle{plainnat}
\bibliography{final_refs}
\end{document}